\documentclass{article}
\usepackage{spconf,amsmath,graphicx}
\usepackage{hyperref}
\usepackage{multirow, booktabs}
\usepackage{enumitem}
\usepackage{pifont}

\title{Semi-Supervised Singing Voice Separation with Noisy Self-Training}
\name{Zhepei Wang$^{\sharp}$ \sthanks{Work performed while at Amazon Web Services.}, Ritwik Giri$^{\dagger}$, Umut Isik$^{\dagger}$, Jean-Marc Valin$^{\dagger}$, Arvindh Krishnaswamy$^{\dagger}$ }
\address{$^{\dagger}$Amazon Web Services\\
$^{\sharp}$University of Illinois at Urbana-Champaign
}
\begin{document}
\ninept
\maketitle
\begin{abstract}
Recent progress in singing voice separation has primarily focused on supervised deep learning methods. However, the scarcity of ground-truth data with clean musical sources has been a problem for long. Given a limited set of labeled data, we present a method to leverage a large volume of unlabeled data to improve the model's performance. Following the noisy self-training framework, we first train a teacher network on the small labeled dataset and infer pseudo-labels from the large corpus of unlabeled mixtures. Then, a larger student network is trained on combined ground-truth and self-labeled datasets. Empirical results show that the proposed self-training scheme, along with data augmentation methods, effectively leverage the large unlabeled corpus and obtain superior performance compared to supervised methods.
\end{abstract}
\begin{keywords}
Singing voice separation, self-training, self attention, data augmentation
\end{keywords}
\section{Introduction}
\label{sec:intro}

The task of singing voice separation is to separate the input mixture into different components: singing voice and accompaniment. It is a crucial problem in music information retrieval and has commercial usage such as music remixing and karaoke applications. It also has the potential to provide useful information for downstream tasks such as song identification, lyric transcription, singing voice synthesis, and voice cloning without access to clean sources.


Deep learning models have recently shown promising results in singing voice separation. Popular methods are mostly supervised methods, where a deep neural network is trained on a multi-track corpus with paired vocal and accompaniment ground-truths.
\cite{mdn, mmdenselstm} apply dense connections between convolutional or long short-term memory (LSTM) blocks to estimate separate masks, and \cite{openunmix} uses a bidirectional LSTM (BLSTM) network in the separator. Models with multi-scale processing further improve the performance of separation. With the concatenation of features at different scales along with skip connections, U-Net \cite{WaveUNetAM} can maintain long term temporal correlation while processing local information with higher resolution. Such architecture has been effective in both time-frequency domain \cite{spleeter2020, spleeter_study_data, M-UNet, attn_unet} and end-to-end, time-domain methods \cite{demucs1, demucs2}.
Models that simultaneously process features at different resolutions with multiple paths have also shown effectiveness in singing voice separation systems \cite{mulcat, meta-tasnet}.

The primary challenge for supervised methods with deep learning is the lack of training data with ground-truth. It is more significant for larger networks that are more prone to overfitting issues. There are several multi-track datasets publicly available for singing voice separation including MIR-1K \cite{mir-1k}, ccMixter \cite{ccmixter}, and MUSDB \cite{musdb18}. However, these datasets are relatively small (all these combined are around 15 hours) and not diverse. To artificially increase the size of the dataset, \cite{spleeter_study_data, aug_uhlich, aug_cohen} apply data augmentation to signal including random channel swapping, amplitude scaling, remixing sources from different songs, time-stretching, pitch shifting, and filtering. These methods, individually or combined, are empirically shown to enhance separation performance only by a limited margin \cite{spleeter_study_data}.

On the other hand, semi-supervised and unsupervised methods do not require a large corpus with a one-to-one correspondence between the mixtures and ground-truth sources. \cite{demucs1} leverages mixture data by first training a silent-source detector on a small labeled dataset, then mixing recordings with only one source and mixture recordings with the source being silent, and finally optimizing with a weakly supervised loss. \cite{adv_stoller, adv_mich} propose generative adversarial frameworks that require isolated sources only. The distance between the distributions of separator's output and the isolated sources is minimized with adversarial training. \cite{dae_1, dae_sinkhorn} use unpaired vocal and accompaniment data to learn non-negative, smooth representations with a denoising auto-encoder using an unsupervised objective. \cite{BootstrappingDM} proposes a stage-wise algorithm where a clustering-based labeler assigns time-frequency bin labels with confidence measure, and a student separator network is trained on these labels afterward.

Self-training is a semi-supervised framework in which a pre-trained teacher model assigns pseudo-labels for unlabeled data. Then a student model is trained with the self-labeled dataset. It has been applied in several applications such as image recognition \cite{selftrn_rec_noisy_2019} and automatic speech recognition \cite{selftrn_asr_noisy_2020}. Our approach follows the noisy self-training method in which we investigate data augmentation methods on musical signal and evaluate how they affect the separation performance.

With the framework of noisy self-training, we aim to improve the performance of a deep separator network where only a limited amount of data with ground-truth is available. The contribution of this work is listed as follows:
\begin{itemize}
    \item We use a large unlabeled corpus to improve separation results under the noisy self-training framework.
    
    \item We show how data augmentation can improve the model's ability to generalize with a focus on random remixing between sources.
    
    \item We propose to use a voice activity detector to evaluate the quality of self-labeled data in the student training to perform data filtering.
\end{itemize}

\section{System Description}
\label{sec:sys}

\subsection{Noisy Self-Training for Singing Voice Separation}
\label{ssec:sys_nst}
Our proposed self-training framework for singing voice separation consists of the following steps:
\begin{enumerate}
    \item Train a teacher separator network $\mathcal{M}_0$ on a small labeled dataset $\mathcal{D}_l$.
    \item Assign pseudo-labels for the large unlabeled dataset $\mathcal{D}_u$ with $\mathcal{M}_0$ to obtain the self-labeled dataset $\mathcal{D}_{0}$.
    \item Filter data samples from $\mathcal{D}_{0}$ to obtain $\mathcal{D}_{f0}$.
    \item Train a student network $\mathcal{M}_1$ with $\mathcal{D}_l \cup \mathcal{D}_{f0}$. 

\end{enumerate}

\begin{figure}[htb]

\begin{minipage}[b]{1.0\linewidth}
  \centering
  \centerline{\includegraphics[width=\textwidth]{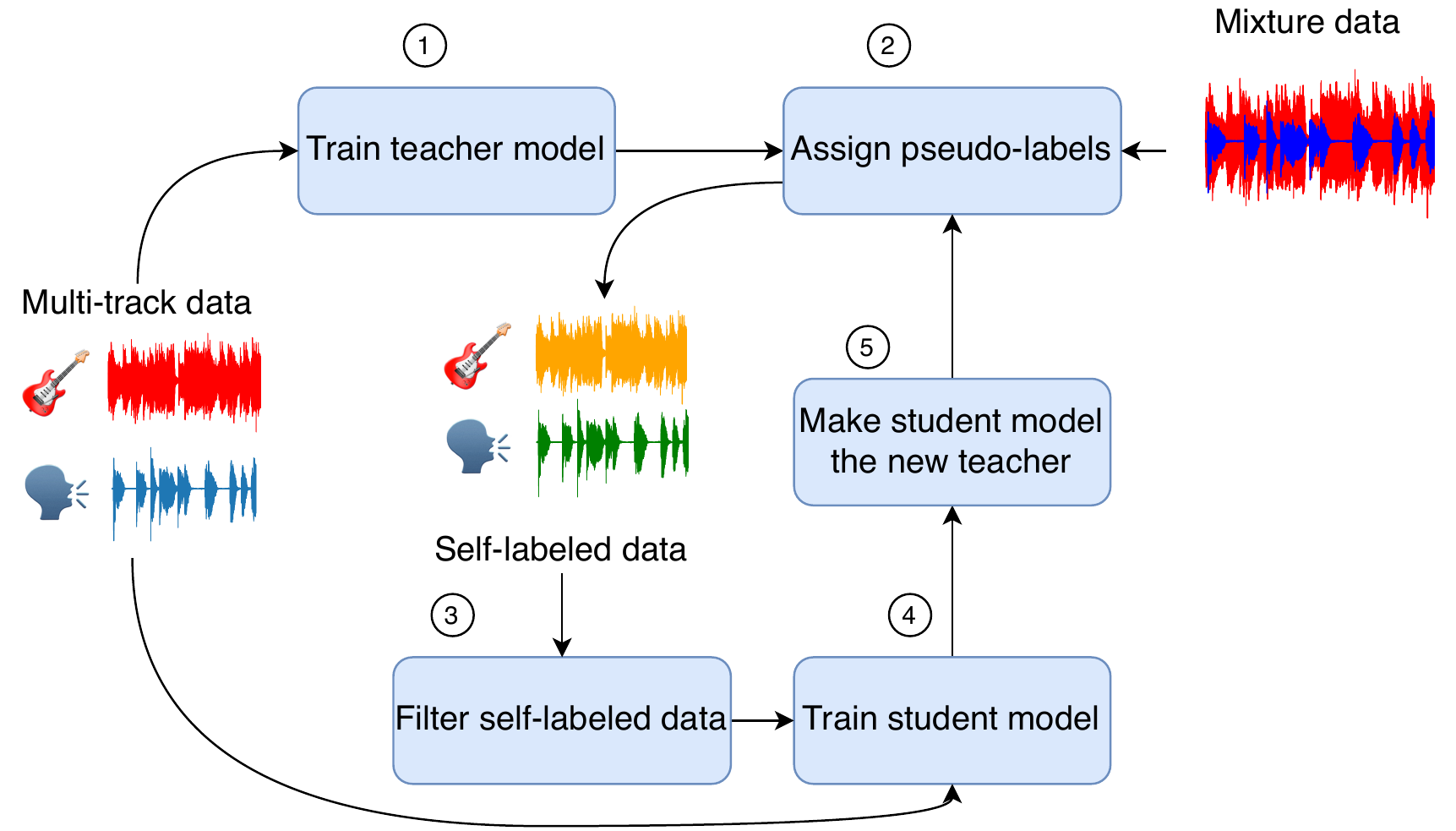}}
\end{minipage}

\caption{The pipeline of noisy self-training for singing voice separation.}
\label{fig:selftrain}
\end{figure}

    This framework can be made iterative by repeating steps 2 to 4, using the student network $\mathcal{M}_i$ as the new teacher to obtain a self-labeled dataset $\mathcal{D}_{i+1}$, and training a new student model $\mathcal{M}_{i+1}$. The process stops when there is no performance gain. We illustrate the framework pipeline in Figure~\ref{fig:selftrain}.
    
\subsubsection{Data Filtering with Voice Activity Detector}
\label{sssec:vad_filter}

Poor quality self-labeled samples may contain leakage of the singing voice in the accompaniment tracks or leakage of musical background in the vocal tracks. To filter out these samples, we evaluate the quality of the data with a voice activity detector (VAD).

The VAD takes the STFT magnitude spectrogram of the mixture as input and predicts the frame-level energy ratio between the source and the mixture. We use the 2D-CRNN architecture with the same configuration as in \cite{vad_univ_sep}. We train two separate VADs to estimate the energy ratio of vocal over mixture, and accompaniment over mixture, respectively. The ground-truth is defined as 0 when both the vocal and the accompaniment are silent. The VADs are trained with the same labeled dataset as the one for the teacher separator model using binary cross-entropy loss.

To measure the leakage of accompaniment in vocal, we pass the self-labeled vocal track into the accompaniment activity detector. Similarly, we feed the self-labeled background track into the vocal activity detector to detect leakage of the singing voice. A frame is defined as a ``poor quality frame'' if either its accompaniment energy in the vocal track or its vocal energy in the background track is higher than some threshold. We count the total number of ``poor quality frames'' for each song, and songs with a smaller percentage of such frames are considered to have higher quality.

\subsubsection{Data Augmentation}
\label{sssec:data_aug}

Data noise is a key component in the noisy self-training framework. We apply data augmentation methods for the training of both teacher and student models. Each training sample contains both vocal and accompaniment tracks of duration 30 seconds. To augment the training set, we randomly select a window of duration $T$ seconds (with $T < 30$) from the sample. We also perform ``random mixing'' by mixing vocal and background sources from two randomly selected songs with a probability of $p$. Besides, we apply dynamic mixing ratio, pitch shifting, lowpass filtering, and EQ filtering to the data.

\subsection{Separator Network}
\label{ssec:sep_network}
We use the PoCoNet \cite{poconet} for both teacher and student models. The neural network takes the concatenation of real and imaginary parts of the mixture's STFT spectrogram as input. The separator estimates the complex ratio masks for each source. The wave-form signal is obtained by applying inverse STFT transform on the estimated spectrograms.

The separator is a fully-convolutional 2D U-Net architecture with DenseNet and attention blocks. Each DenseNet block contains three convolutional layers, each followed by batch normalization and Rectified Linear Unit (ReLU). Convolutional operations are causal only in the time direction but not in the frequency direction. We choose a kernel size of $3\times 3$ and a stride size of 1, and the number of channels increases from 32, 64, 128 to 256. We control the size of the network by varying the number of levels in U-Net and the maximum number of channels. In the attention module, the number of channels is set to 5 and the encoding dimension for key and query is 20. The connections of layers in DenseNet and attention blocks follow \cite{poconet}. Frequency-positional embeddings are applied to each time-frequency bin of the input spectrogram. For time frame $t$ and frequency bin $f$, the embedding vector is defined as:
\begin{equation}
    \rho(t, f) = (\cos(\pi \frac{f}{F}), \cos(2\pi \frac{f}{F}), \ldots, \cos(2^{k-1}\pi \frac{f}{F})),
    \label{eq:pos_emb}
\end{equation}
where $F$ is the frequency bandwidth and $k = 10$ is the dimension of the embedding.




\subsubsection{Loss Functions}
\label{sssec:sys_loss}

For each output source, the loss function is the weighted sum of wave-form and spectral loss:
\begin{equation}
    \mathcal{L}_s(y, \hat{y}) = \lambda_{\text{audio}}\mathcal{L}_{\text{audio}}(y, \hat{y}) + \lambda_{\text{spec}}\mathcal{L}_{\text{spec}}(Y, \hat{Y}),
    \label{eq:loss_source}
\end{equation}
where $s$ is the output source, $y, \hat{y}$ are time-domain output and reference signals and $Y = \lvert \text{STFT}(y)\rvert, \hat{Y} = \lvert \text{STFT}(\hat{y})\rvert$ are the corresponding STFT magnitude spectrograms. We choose both $\mathcal{L}_{\text{audio}}(\cdot)$ and $ \mathcal{L}_{\text{spectral}}(\cdot)$ to be $\ell1$ loss. The total loss is the weighted sum of each source:
\begin{equation}
    \mathcal{L}(y, \hat{y}) = \lambda_{\text{voc}}\mathcal{L}_{\text{voc}}(y, \hat{y}) + \lambda_{\text{acc}}\mathcal{L}_{\text{acc}}(Y, \hat{Y}),
    \label{eq:loss_total}
\end{equation}
\section{Experimental Setup}
\label{sec:expr}

\subsection{Dataset}
\label{ssec:expr_dataset}

We use MIR-1K \cite{mir-1k}, ccMixter \cite{ccmixter}, and the training partition of MUSDB \cite{musdb18} as the labeled dataset for supervised training. The training set contains approximately 11 hours of recordings. We use DAMP \cite{damp} as the unlabeled dataset for training the student model. DAMP dataset contains more than 300 hours of vocal and background recordings from karaoke app users. Since these recordings are not professionally produced, there exists bleeding of music in the vocal tracks and bleeding of singing voice in the accompaniment tracks as well; hence, it is not suitable for supervised source separation.

\subsection{Preprocessing}
\label{ssec:expr_preprocess}
To reduce dimensionality and speed up processing, we downsample each track to 16 kHz and convert it to mono. We further segment the recordings to non-overlapping 30-second segments, and if the segment is less than 30 seconds we zero-pad to the end of the signal. The spectrograms are computed with a 1024-point STFT with a hop size of 256.

\subsection{Noisy Self-Training Procedure}
\label{ssec:expr_trn}
For both teacher and student training, we minimize Equation~\ref{eq:loss_total} with an Adam optimizer with an initial learning rate of 1e-4, and we decrease the learning rate by half for every 100k iterations until it's no greater than 1e-6. We set $\lambda_{\text{audio}} = \lambda_{\text{spec}} = 1$ in Equation~\ref{eq:loss_source} and $\lambda_{\text{voc}} = \lambda_{\text{acc}} = 1$ in Equation~\ref{eq:loss_total}.

To augment the training set, we randomly select a window size of $T = 2.5, 5, 10$ seconds as the input to the model to experiment with the effect of input length, each with a batch size of 4, 2, and 1, respectively. The maximal batch size is chosen under the memory limit. We experiment with different probabilities of applying random mixing with $p = 0, 0.25, 0.5, 0.75, 1$.

The teacher model is trained on the labeled datasets. Then, it assigns pseudo-labels for the unlabeled dataset. We infer vocal labels using DAMP vocal tracks as input to the teacher model and infer accompaniment labels from DAMP accompaniment tracks. Due to the leakage in these vocal and background tracks, they can be viewed as mixtures where one source is more likely to dominate the other, compared to normal mixtures.

\section{Evaluation Results and Discussions}
\label{sec:res}

\begin{table}[ht]
\centering
\begin{tabular}{c|c|c|c|c|c|c}
\hline
Len & Size & Prob & Use & SDR(V) & SDR(A) & Mean \\
(s) & (1e6) & RM & DAMP & & & \\
\hline
\multirow{2}{*}{2.5} & \multirow{10}{*}{8.3} & 0 & \multirow{4}{*}{No} & 1.84 & 10.31 & 6.08 \\
& & 0.5 & & 1.72 & 9.51 & 5.62 \\
 \cmidrule{1-1}\morecmidrules\cmidrule{3-3}\morecmidrules\cmidrule{5-7}
\multirow{2}{*}{5} & & 0 & & 3.55 & 10.91 & 7.23 \\
& & 0.5 & & 4.08 & 11.34 & 7.71 \\
 \cmidrule{1-1}\morecmidrules\cmidrule{3-7}
\multirow{11}{*}{10} & & 0 & Yes & 3.93 & 11.46 & 7.70 \\
\cmidrule{3-7}
& & 0 &  \multirow{5}{*}{No} & 5.88 & 12.52 & 9.2 \\
& & 0.25 & & 6.35 & 12.56 & 9.46 \\
& & 0.5 & & 7.06 & 13.35 & 10.21 \\
& & 0.75 & & 6.98 & 13.36 & 10.17 \\
& & 1.0 & & 6.91 & 13.66 & \textbf{10.29} \\
\cmidrule{2-7}
& \multirow{3}{*}{1.6} & 0 & Yes & 0.03 & 6.62 & 3.33 \\
\cmidrule{3-7}
&  & 0 &  \multirow{2}{*}{No} & 4.17 & 10.86 & 7.52 \\
& & 0.5 & & 4.34 & 11.13 & 7.74 \\
\cmidrule{2-7}
& \multirow{2}{*}{15.4} & 0 & \multirow{2}{*}{No} & 5.81 & 11.94 & 8.88 \\
& & 0.5 & & 6.9 & 13.07 & 9.99 \\

\hline
\end{tabular}
\caption{Test performance metrics (SDR in dB) for teacher model candidates. We experiment with various input sizes, number of model parameters, and the probability of random mixing to pick the best configuration for the teacher model. The best performance is highlighted in bold.}
\label{tab:teacher}
\end{table}

\begin{table}[!htb]
\centering
\begin{tabular}{c|c|c|c|c}
\hline
Size & top \% & SDR(V) & SDR(A) & Mean \\
(1e6) &  & &  &  \\
\hline
8.3 & 1 & 6.57 & 12.92 & 9.75 \\
\hline
\multirow{3}{*}{15.4} & 1 & 7.27 & 13.73 & 10.5 \\
& 0.5 & 7.52 & 13.91 & 10.72 \\
& 0.25 & 7.8 & 13.92 & \textbf{10.86} \\
\hline
\end{tabular}
\caption{Test performance metrics (SDR in dB) for student models. We experiment with different model sizes and the proportion of quality-controlled self-labeled samples. The best performance is shown in bold.}
\label{tab:student}
\end{table}

\begin{table*}[htb]
\centering
\begin{tabular}{c|c|c|c|c|c|c}
\toprule
Name & \#Src & Input & Extra & SDR(V) & SDR(A) & Mean \\
& & type & Data & & & \\
\hline \hline
Demucs\cite{demucs2} & \multirow{2}{*}{4} & \multirow{2}{*}{Stereo} & Labeled & 7.05 & N/A & N/A \\
\cite{mulcat} &  &  & \ding{55} & 6.92 & N/A & N/A \\ 
\hline
MMDenseLSTM\cite{mmdenselstm} & \multirow{4}{*}{2} & \multirow{2}{*}{Stereo} & \ding{55} & 4.94 & \textbf{16.4} & 10.67 \\
MDN\cite{mdn} &  &  & \ding{55} & 3.87 & 15.41 & 9.64 \\
\cmidrule{1-1}\morecmidrules\cmidrule{3-7}
MT U-Net\cite{M-UNet} &  & \multirow{2}{*}{Mono} & \ding{55} & 5.28 & 13.04 & 9.16 \\ 
\cite{adv_mich} &  & & \ding{55} & 3.5 & N/A & N/A \\
\midrule
Ours (teacher) & \multirow{3}{*}{2} & \multirow{3}{*}{Mono} & \ding{55} & 6.91 & 13.66 & 10.29 \\
Ours (student, no VAD) & & & DAMP & 7.27 & 13.73 & 10.5 \\
Ours (student, VAD) & & & DAMP & \textbf{7.8} & 13.92 & \textbf{10.86} \\
\bottomrule
\end{tabular}
\caption{Comparison of the proposed method and other baseline models. The best performance is shown in bold.}
\label{tab:comp}
\end{table*}

\subsection{Evaluation Framework}
\label{ssec:res_eval}
As in previous studies on singing voice separation \cite{mdn, mmdenselstm, WaveUNetAM, M-UNet, adv_mich}, we measure the signal-to-distortion ratio (SDR) to evaluate the separation performance. Following the SiSec separation campaign \cite{sisec}, we use the 50 songs from the test partition of MUSDB \cite{musdb18} as the test set. We partition each audio track into non-overlapping one-second segments, and we take the median of segment-wise SDR for each song and report the median from all 50 songs. We use the python package \texttt{museval}\footnote{\url{https://sigsep.github.io/sigsep-mus-eval/}} to compute SDR.

\subsection{Teacher Training}
\label{ssec:res_teacher}


We select the configuration for the teacher model by experimenting with different input window sizes of training samples and the number of model parameters. Table~\ref{tab:teacher} shows the test SDR for the combinations of input and model size. We first observe that using longer input size improves both vocal and accompaniment SDR. The improvement can be attributed to the attention blocks where longer input context provides more information for separation. Another observation is that larger models do not guarantee performance gain. The largest model (15.4M param) performs significantly better than the smallest one (1.6M) but is slightly worse than the 8.3M version for the probability of random mixing $p=0, 0.5$. Using the best combination of input length (10 seconds) and model size (8.3M), we experiment with different probability of applying random mixing. \cite{spleeter_study_data} shows that random mixing does not have a positive effect on test SDR, and one possible explanation is that it creates mixtures with somewhat independent sources. Our experiments, however, indicate that random mixing alone significantly improves the results. The best performance is obtained when random mixing is always applied. Our observations are consistent with the argument in \cite{ismir2020_tutorial} that ``one-versus-all'' separation benefits from mixing independent tracks.
Intuitively, mixtures with dependent sources are more difficult to separate. Random mixing makes it easier for the model to learn and to converge faster on the training set. Meanwhile, by mixing up sources from different songs, the training set becomes more diverse and the model has a better ability to generalize at inference time. 

In addition, we verify that the DAMP dataset should not be applied directly in supervised source separation tasks by including this dataset along with the other labeled datasets. We experiment with two model sizes (1.6M and 8.3M) using 10-second input without random mixing, and the SDR values degrade sharply for both cases.


\subsection{Student Training}
\label{ssec:res_student}


Table~\ref{tab:student} summarizes the test SDR for student models. As opposed to the teacher model, the 15.4M model has a 0.75 dB SDR gain compared to the 8.3M model. The observation that the larger capacity student model improves the performance is consistent with the findings in \cite{selftrn_rec_noisy_2019}.

\begin{figure}[htb]
\begin{minipage}[b]{0.9\linewidth}
  \centering
  \centerline{\includegraphics[width=\textwidth]{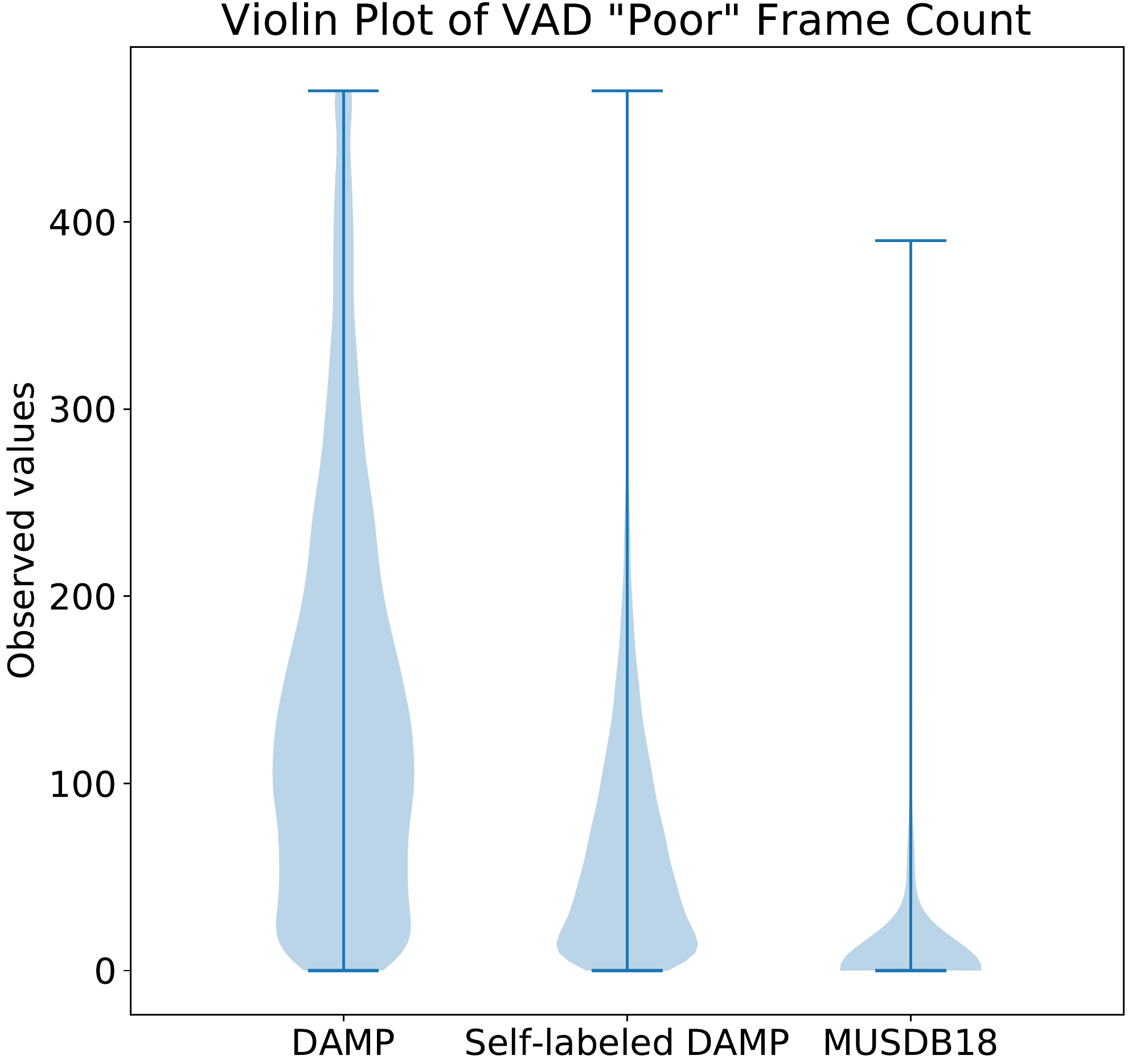}}
\end{minipage}

\caption{Count of ``poor-quality frames'' for different datasets.}
\label{fig:vad_cnt}
\end{figure}

To verify the quality control approach with VADs, we first count for each song the number of ``poor-quality frames'' as defined in Section~\ref{sssec:vad_filter} from three different datasets: DAMP, self-labeled DAMP, and MUSDB. From the visualization in Figure~\ref{fig:vad_cnt}, the unprocessed DAMP contains the highest percentage of data with a large number of poor-quality frames, the distribution of MUSDB is concentrated in the low count region, while the self-labeled dataset lies in between. This implies that the count of ``poor-quality frames'' based on the output of VADs is a reasonable indicator of the quality of data samples. The experimental results demonstrate that the proposed data filtering method with VADs further improves the performance. The highest SDR is obtained when only the top quarter of the self-labeled data is included in the training.
Incorporating a higher percentage of self-labeled data may provide more diversity but is more likely to include samples with poor quality, thus negatively affecting the model's performance. 

\subsection{Comparison with Other Methods}
\label{ssec:res_comp}
To compare the separation of singing voice with state-of-the-art, we also include models that separate the mixture into four sources. It has been shown in \cite{spleeter_study_data} that, these four-source models have similar vocal separation performance compared to two-source models, even though the four-source separation task is more challenging than the two-source counterpart; possibly because of the additional supervision provided by different instrumental sources in the multi-task learning setup. Hence, we include the vocal SDR values of state-of-the-arts for four-source models \cite{demucs2,mulcat} in our comparison. 
Our proposed approach, the student model using quality control with VADs, obtains the highest vocal and average SDR among all models, and the vocal separation outperforms others by a significant margin. The accompaniment SDR is higher than the baseline model with mono input \cite{M-UNet} but worse than the stereo ones \cite{mdn, mmdenselstm}. Stereo input contains more spatial information for accompaniment than vocal since the left and right channel difference for background tracks are at a much larger scale than vocal tracks. Such information may improve the separation of accompaniment.

\section{Conclusion}
\label{sec:conc}

We present a semi-supervised method for singing voice separation to deal with the scarcity of data with ground-truth. Using the noisy self-training framework, we can effectively make use of a large unlabeled dataset to train a deep separation network. Experimental results show that random mixing as data augmentation improves model training, and the data filtering method with pre-trained voice activity detectors improves the quality of the self-labeled training samples. Our study serves as a foundation for more complicated systems such as using stereo input, working with unlabeled datasets with mixture only (as opposed to noisy source tracks), and extending the teacher-student loop with additional iterations.

\bibliographystyle{IEEEbib}
\bibliography{refs}

\end{document}